\documentclass[12pt, a4paper,preprint]{article}
\usepackage{jheppub}
\usepackage{bm} 
\usepackage{tikz}

\def\nn{\nonumber}
\def\bea{\begin{eqnarray}}
\def\eea{\end{eqnarray}}
\def\beas{\begin{eqnarray*}}
\def\eeas{\end{eqnarray*}}
\def\mau{\mathcal{U}}
\def\mf{\mathcal{F}}
\title{Spherical Contours, IR divergences and the geometry of Feynman parameter integrands at one loop}
\author{Akshay Yelleshpur Srikant}
\affiliation[b]{Department of Physics, Princeton University, NJ, USA}
\abstract{Spherical contours introduced in \cite{SphericalContours} translate the concept of ``discontinuity across a branch cut" to Feynman parameter space. In this paper, we further explore spherical contours and connect them to the computation of leading IR divergences of 1 loop graphs directly in Feynman parameter space. These spherical contours can be used to develop a Feynman parameter space analog of ``Leading Singularities" of loop integrands which allows us to develop a method of determining Feynman parameter integrands with no reference to the momentum space loop integrand. Finally, we explore some interesting features of Feynman parameter integrands in $\mathcal{N}=4$ SYM.}
\begin{document}
\maketitle
\newpage
\section{Introduction}
\label{sec:intro}
A connection between the singularity structure of one-loop integrands and the projective geometry of their associated Feynman parameter integrand was established in \cite{SphericalContours}. One of the central results of this paper was the introduction of a new kind of residue in Feynman parameter space - associated with ``spherical contours " - which capture information about discontinuities of the integrands across various branch cuts. It was also shown that this seemingly calculus based operation also has an algebraic interpretation. The purpose of this paper is to provide some additional details on this algebraic interpretation and also explore IR divergent integrals as the authors of \cite{SphericalContours} largely focused on finite integrals. \\

The structure of the paper is as follows. We begin by discussing some preliminaries of Feynman parametrization and setting the notation for the rest of the paper in Section [\ref{sec:FPrevisit}]. In Section [\ref{sec:IRdiv}], we investigate IR divergent integrals in Feynman parameter space. We motivate and develop a new kind of ``residue" operation which computes the leading IR divergence of one loop amplitudes and show that it correctly reproduces the leading IR divergences in all cases. Section [\ref{sec:aspects}] involves discussion of the algebraic structure of spherical residues. A method to construct one loop integrands using spherical residues is outlined in Section [\ref{sec:construct}]. We conclude by examining some appealing features of Feynman parameter integrands in $\mathcal{N}=4$ SYM in [\ref{sec:oneloopfeynpar}]. 

\section{Feynman Parametrization revisited}
\label{sec:FPrevisit}
Although Feynman parametrization is a familiar trick, let us begin by discussing it in a more geometric way. This will highlight some of the features of Feynman parameter integrals which are important for the rest of the paper. consider the scalar one-loop integrals of the form ($\mu^2$ is the mass scale introduced in dimensional regularization)
\bea
\label{eq:scalarintegral}
I_n = (\mu^2)^{\nu - \ell D/2} \int \prod_{k=1}^L \frac{d^D\ell_k}{i\pi^{D/2}}\prod_{j=1}^n \frac{1}{(-q_j^2+m_j^2)^{\nu_j}} \qquad \nu = \sum_{j=1}^n \nu_j
\eea
Here each $q_j$ is a linear combination of the external momenta $p_k$ and the loop momenta $\ell_k$. A straightforward Feynman parametrization yields
\bea
\label{eq:origfeynpar}
I_n = (\mu^2)^{\nu - \ell D/2} \frac{\Gamma(\nu)}{\prod_{j=1}^n \Gamma(\nu_j)}\int_0^\infty d^n \,x\, \delta(1-\sum_i x_i)\prod_{j=1}^n x_j^{\nu_j -1}\frac{\mathcal{U}^{n-4}}{\mathcal{F}^{n-2}}
\eea
where $\mathcal{U}$ and $\mathcal{F}$ are functions of the Feynman parameters $x_j$ which depend on the particular integral. They are defined by first expressing the denominator as a polynomial in the loop momenta.
\beas
\sum_{j=1}^n x_j (-q_j^2+m_j^2) = -\sum_{r,s=1}^{L} \ell^\mu_r M_{r,s} \ell_{s\mu} + 2 \sum_{r=1}^L l_r^\mu Q_{r\mu} + J
\eeas
where $J$ contains all the terms independent of the loop momenta. Then, 
\beas
\mathcal{U} = \text{det}M \qquad\qquad \mathcal{F}=\text{det}M ( J + Q.M^{-1}.Q)
\eeas
$\mau$ and $\mf$ are called the Symanzik polynomials. It will be of interest to note that for one loop $\mathcal{U}$ and $\mathcal{F}$ are homogenous polynomials which are linear and quadratic respectively. They can also be calculated efficiently by using graphical rules which are detailed in \cite{symanzik}.\\

It is illuminating to consider another path of arriving at this result. Let us first introduce Schwinger parameters $\alpha_i$ 
\bea
\label{eq:schwinger}
\frac{1}{(-q_i^2+m_i^2)^{\nu_i}} = \int_0^\infty d \alpha_i \, e^{-\alpha_i (-q_i^2+m_i^2)^{\nu_i}}
\eea
Inserting this in [\ref{eq:scalarintegral}], we can perform the Gaussian integrals over all the loop momenta. The result is
\bea
\label{eq:alpharep}
I_n = (\mu^2)^{\nu - \ell D/2} \frac{i^{-\nu -1}\pi^2}{\prod_{j=1}^n \Gamma(\nu_j)} \int_0^\infty d\alpha_1 \dots d\alpha_n \, \prod_{j=1}^n \alpha_j^{\nu_j -1} \frac{1}{U^2} e^{i \frac{F}{U} - i\sum_j m_j^2 \alpha_j}
\eea
where $U$ and $F$ are polynomials in the $\alpha_i$. They are homogenous and like the Symanzik polynomials $\mau$ and $\mf$, linear and quadratic respectively. For more details, see \cite{symanzik, smirnov}.\\

We can now introduce new variables via $\alpha_i =\eta x_i$. Since there are $n+1$ new variables, we must impose a constraint on the $x_i$ which we take to be $\sum_{i\in S} x_i = 1$ where $S \subset \left\lbrace 1, \dots n\right\rbrace$ which changes [\ref{eq:alpharep}] to
\beas
I_n = (\mu^2)^{\nu - \ell D/2} \frac{i^{-\nu -1}\pi^2}{\prod_{j=1}^n \Gamma(\nu_j)} \int_0^\infty dX\,d\eta\, \eta^{n-1}\, \left( \eta^{\nu-n-2} \, \prod_{j=1}^n x_j^{\nu_j -1} \frac{1}{U^2} e^{i \,\eta \,\left( \frac{F}{U} - i\sum_j m_j^2 \,x_j\right)}\right)
\eeas
where $dX = dx_1 \dots dx_n \,\delta (1 - \sum_{i \in S} x_i)$. This result is called the Cheng-Wu theorem \cite{ChengWu}. In particular, this implies that we could set any one of the Feynman parameters $x_i$ to 1. The vector $X = (x_1, \dots , x_n)$ can be thought of as a point in projective space and the measure $dx_1 \dots dx_n \,\delta (1 - \sum_{i \in S} x_i)$ can be better written as
\beas
\langle Xd^{n-1}X\rangle \equiv \epsilon_{\mu_1 \dots \mu_n} X^{\mu_1} \,dX^{\mu_2}\dots dX^{\mu_{n}}.
\eeas
The textbook result of Feynman parametrization [\ref{eq:origfeynpar}] is obtained by setting \\$S = \left\lbrace 1, \dots n\right\rbrace$. For the rest of the paper, we will assume that the factors $\nu_i = 1$ and write all the Feynman parameter integrals in a projective manner as shown below. 
\beas
\label{eq:scalarfint}
I_n = \int \frac{\langle Xd^{n-1}X\rangle \, \mathcal{U}^{n-4}}{\mathcal{F}^{n-2}}
\eeas
The homogeneity properties of $\mathcal{U}$ and $\mathcal{F}$ are essential in making the integrals projectively well defined. \\

Throughout this paper we will use three kinds of variables to describe the external momenta - dual momenta, momentum twistors and embedding space momenta. Dual momenta $y_i^\mu$ are defined by  
\beas
p_i^\mu = y_i^\mu - y_{i-1}^\mu \qquad y^\mu_{ij} \equiv y^\mu_i - y^\mu_j.
\eeas
We associate a variable $y$ for the loop momentum. Momentum twistors $Z_i$ are defined by associating a line $Z_{i-1}Z_i$ with each $y_i$. The scalar $y_{ij}^2$ is related to the $SL(4,R)$ invariant $\langle i-1ij-1j\rangle \equiv \epsilon_{ABCD}Z_{i-1}^A Z_i^B Z_{j-1}^C Z_j^D$. Each loop momentum variable is associated to a line $AB$ in twistor space which is to be integrated over using the measure $\langle ABd^2A \rangle \langle ABd^2B\rangle$. For more details, see \cite{LocalIntegrands}\\

A vector $y^\mu$ in D-dimensional Minkowski space is mapped to a null vector $Y^M = (1, y^2, y^\mu)$ in embedding space. Here, we have specified the components in light-cone co-ordinates, i.e. $Y^+ = 1,Y^- = y^2$ and $Y^\mu = y^\mu $. The metric is $g_{+-}=g_{-+} = -1/2$ and $g_{\mu \nu}=\eta_{\mu\nu}$ with all other entries zero. The invariants $y_{ij}^2 = -2Y_i. Y_j $. In particular, for null momenta, we have $Y_i.Y_{i+1}=0$. The integral $(\ref{eq:scalarintegral})$ can be written as
\bea
\label{eq:embeddingscalar}
I_n = (\mu^2)^{\nu - \ell D/2} \int \frac{\left[d^4 Y\right]}{(Y.Y_1)^{\nu_1}\dots (Y.Y_n)^{\nu_n}}
\eea
The measure $[d^4Y] = \frac{d^6Y \delta(Y.Y)}{\text{Vol}(GL(1))}$. For more details, see \cite{SphericalContours, embedding}.\\

For the particular case of planar one-loop integrals, simple expressions are available for the Symanzik polynomials. While $\mau$ depends on the details of the numerator, $\mf$ depends only on the pole structure.
\beas
\mathcal{F}=\sum_{i<j}x_i x_j \, y_{ij}^2 = X.Q.X 
\eeas
where $Q_{ij}$ can be expressed in any of the three equivalent forms $y_{ij}^2$ ,$\langle i-1ij-1j \rangle$ or $Y_i.Y_j$.

\section{1-loop IR divergences}
\label{sec:IRdiv}
It is well known that loop integrals suffer from IR divergences. These divergences arise when the loop momentum $\ell$ becomes collinear with an external massless momentum $p_i$, i.e. $\ell.p_i \rightarrow 0$ (soft) or when it becomes collinear to two consecutive null external momenta $\ell.p_{i-1} = \ell.p_{i} = 0$. As a specific example, consider the 4D massless box integral in momentum space and the corresponding Feynman parameter integral. 
\begin{figure}[htb!]
\begin{center}
\includegraphics[scale=.5]{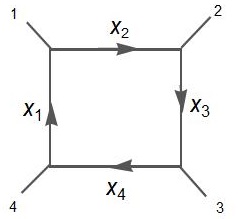}
\end{center}
\end{figure}
\beas
I_4  = \int d^4\ell\, \frac{1}{\ell^2 \,(\ell - p_2)^2\, (\ell - p_2 - p_3)^2 \, (\ell + p_1)^2 } \, = \, \int \frac{\prod_{i=1}^4 dx_i \, \delta(x_1 + x_2+x_3+x_4-1)}{(x_1 x_3\, s + x_2 x_4\, u)^2}
\eeas
where $s = (p_1 + p_2)^2$ and $u = (p_1+p_4)^2$. This integral is of course well known and has been evaluated in dimensional regularization in $D = 4-2\epsilon$ dimensions.
\beas
I_4 = \frac{\Gamma(1+\epsilon) \Gamma^2(1-\epsilon)}{\Gamma(1-2\epsilon)s \, u}\left( \frac{2}{\epsilon^2}\left[ (-\mu^{-2} s)^{-\epsilon} + (-\mu^{-2} u)^{-\epsilon}\right] - \text{log}^2 \frac{s}{u} - \pi^2 \right) + \mathcal{O}(\epsilon)
\eeas
The presence of the $\frac{1}{\epsilon^2}$ terms indicated an IR divergence. A more transparent analysis using a massive regulator instead of dimensional regularization reveals that the region divergence coming from the regions $\ell^2 = (\ell - p_2)^2 = (\ell + p_1)^2 = 0$ is of the form $\frac{1}{s u}( \text{log}^2 \frac{m^2}{s} + \text{log}^2 \frac{m^2}{u})$. We refer the reader to section 7 in \cite{smatintwistorspace}.\\

We can thus precisely characterize the IR divergent region in momentum space as being associated with three propagators going on-shell. A similar characterization in Feynman parameter space should involve the Feynman parameters corresponding to these three propgators, $x_1, x_2$ and $x_3$. To motivate such a characterization, recall the definition of the Schwinger parameter $\alpha$
\beas
\frac{1}{p^2} = \int_0^\infty d\alpha \, e^{- \alpha \, p^2}
\eeas
Near the upper limit of the integral, i.e. for large $\alpha$, a configuration with $p^2 \approx 0$ would be the most relevant. We might hope that the large $\alpha$ limit probes soft momenta.  Since Feynman parameters are related to Schwinger parameters by $x_i = \frac{\alpha_i}{\sum_i \alpha_i}$, the limit $\alpha_i \rightarrow \infty$ corresponds to $x_i \rightarrow 1$ projectively or $x_i \rightarrow \infty$ non projectively. \\

Furthermore, we can manipulate $I_4$ to understand the relationship between the consecutive massless legs as follows. We use Lorentz invariance to transform to a frame in which $p_1^{\mu} = (1,0,0,0)$ and $p_2^{\mu} = (0,1,0,0)$. Here, we have specified the components in light-cone coordinates as $(p_i^+, p_i^-,p_i^1,p_i^2)$. Hence $p_1^2 = p_2^2 = 0$ is automatic. If we work in the soft region where $\ell^2 \approx 0$, we can write $\ell. p_1 \approx l_+$ and $\ell .p_2 \approx l_-$.
\bea
I_4^M = \int \frac{1}{2 p_2.p_3}\frac{d^4 \ell}{(\ell_+ \ell_- - \ell_\perp^2)\, \ell_- \,  \ell_+}
\eea
The soft collinear region is the region in which all three propagators go on shell. $\ell_+ \approx 0$, $\ell_- \approx 0$ and $\ell_\perp^2 \approx \ell_+ \,\ell_-$. This suggests that the corresponding region in Feynman parameter space is $x_2 \approx x_1 x_3$ and $x_2 \rightarrow \infty$. In this region, we should be able to observe a log$^2$ divergence and calculate its coefficient. This is also equal to the coefficient of the $\frac{1}{\epsilon^2}$ term in dimensional regularization and is known as the cusp anomalous dimension $\Gamma$. In what follows, we will demonstrate that this region in Feynman parameter space indeed captures the IR divergent region and calculates the corresponding  $\Gamma$.
\subsection{Composite residues in momentum space}
\label{sec:momsp}
Let us begin by understanding the calculation of $\Gamma$ directly in momentum space as a composite residue on the poles corresponding to three propagators going on shell. We demonstrate this for the case of a scalar n-gon. 
\bea
\label{eq:scalarint}
I_n && =\nonumber \int \frac{d^4 \,l}{l^2 (l-p_2)^2\,\dots \,(l+p_1)^2} \\
&&=\, \int \frac{d^4y}{(y-y_1)^2(y-y_2)^2\,\dots\, (y-y_n)^2}
\eea
where
\beas
p_i = y_i - y_{i-1} \qquad y=l+x_1 \qquad (l-\sum_{i=2}^k p_i) = l+y_1-y_k
\eeas
We want to calculate the residue associated with the loop momentyum $\ell$ being collinear to two consecutive null external momenta, i.e. $\ell.p_{i-1} = \ell.p_{i} = 0$. In terms of the dual momenta $y_i$, this is equivalent to $(y-y_i)^2 = (y-y_{i-1})^2 = (y-y_{i+1})^2 = 0$. To calculate this residue, we first parametrize $y$ on the cut $(y-y_i)^2 = 0$ by introducing spinor helicity variables. 
\beas
&&y \equiv y_i +\lambda\, \tilde{\lambda} \qquad \qquad y_i - y_{i-1} \equiv \lambda_i \,\tilde{\lambda}_i
\eeas
From this, it follows that
\beas
&&\nn (y-y_{i-1})^2 = 2 \, \left\langle \lambda\,i\right\rangle \,\,\, \left[ \tilde{\lambda} \,\tilde{i+1}\right]\qquad(y-y_{i+1})^2 = 2\,\langle \lambda \,i+1 \rangle \left[ \lambda\, i+1\right] \\
&&\nonumber \,\, (y-y_k)^2 = (y_i-y_k)^2+2  \,(y_i - y_k).(\lambda\tilde{\lambda})
\eeas 
For convenience, we expand $\lambda$ in a basis consisting of $\lambda_i$ and $\lambda_{i+1}$ (with a similar expansion for $\tilde{\lambda}$).
\beas
\lambda =  \beta \lambda_i + \, \gamma\, \lambda_{i+1} \qquad  \tilde{\lambda }= \sigma \, \tilde{\lambda}_i + \,\rho\, \tilde{\lambda}_{i+1}
\eeas
In terms of these variables, the measure
\beas
d^4y = \frac{d^2 \lambda \, d^2\tilde{\lambda}}{\text{Vol }GL(1)} = d\gamma \, d\rho \, d\sigma \, \langle ii+1\rangle\, \left[ii+1\right]
\eeas
We have used the $GL(1)$ to fix $\beta = 1$. By introducing the spinor helicity variables, we are already on the cut $(y-y_i)^2=0$. This residue can now be written as 
\beas
Res_{(y-y_i)^2=0} I_n = \frac{1}{4}\int \frac{d\gamma}{\gamma} \frac{d\rho}{\rho} \frac{d\sigma}{\sigma} \frac{1}{\langle ii+1 \rangle \left[ ii+1\right] \prod_{k\neq \lbrace i-1, i, i+1\rbrace}((y_i - y_k)^2+ 2(y_i-y_k).\lambda \tilde{\lambda})}
\eeas
On this cut, we can now fully localize the loop momentum $y_i$ by taking the residue of the poles $\gamma = \rho = \sigma =0$, even though we have cut only three propagtors. This is an example of a composite residue. Recalling that $\langle ii+1 \rangle \left[ ii+1 \right] = (y_i-y_{i-1}).(y_{i+1}-y_i) =\frac{1}{2} y_{i-1i+1}^2$, the co-efficient of the IR divergence can be written as 
\bea
\label{eq:cusp}
\Gamma_{I_n} = \oint_{\gamma = \rho = \sigma = 0} Res_{(y-y_i)^2=0} I_n = \frac{1}{8}\frac{1}{y_{i-1,i+1}^2\prod_{k\neq \lbrace i-1, i, i+1\rbrace}y_{i,k}^2}
\eea
This can be compared to the full expression for the amplitudes given in \cite{Dimregint, Dimregoneint}.
\subsection{Composite residues in Feynman parameter space}
\label{sec:feynmanres}
We will now demonstrate that the coefficient of the $log^2$ divergence, as obtained in (\ref{eq:cusp}) can also be obtained directly in Feynman parameter space. As suggested above, the IR divergences in Feynman parameter space are associated to a triplet of consecutive Feynman parameters $(x_{i-1},x_i,x_{i+1})$ and come from the region where $x_i$ is large and $x_{i-1}x_{i+1}$ scales as $x_i$. We will evaluate the integral (\ref{eq:scalarint}) in this limit and find that the result is proportional to $\Gamma_{I_n}$. \\
We being by writing (\ref{eq:scalarint}) as a projective integral in Feynman parameter space. 
\bea
\label{eq:scalarfint}
I_n = \int \frac{\langle Xd^{n-1}X\rangle \, \mathcal{U}^{n-4}}{\mathcal{F}^{n-2}}
\eea
with the Symanzik polynomials
\beas
\mathcal{U} = \sum_i x_i \qquad\qquad \mathcal{F}=\sum_{i<j}x_i x_j
\eeas
Let us introduce new variables $(\rho, \tau)$ via
\beas
x_{i-1} = \sqrt{x_i}\, \rho \, e^\tau \qquad x_{i+1}= \, \sqrt{x_i}\, \rho \, e^{-\tau}
\eeas
This change of variables ensures that we have the required scaling, $x_{i-1} x_{i+1} = \rho^2 x_{i}$ of the relevant Feynman parameters. In the limit of the limit of large $x_i$, the Symanzik polynomials reduce to
\beas
&&\mathcal{U} = x_i + \mathcal{O}(\sqrt{x_i})\\
&&\mathcal{F} = x_i \left(y_{i-1i+1}^2 \rho^2 + \sum_{j\neq\lbrace i-1,i,i+1\rbrace}y_{ij}^2\, x_j\right)+ \mathcal{O}(\sqrt{x_i}).
\eeas
Note that the quadric has facotrized in this limit. This guarantees that the resulting integral over the remaining $(n-3)$ Feynman parameters (recall that the integral is projective and requires only $(n-1)$ integrations) is now rational.  
\beas
\nonumber I_n &&\approx \int \prod_{k\neq i-1,i+1}dx_k  \,2\rho\,x_i\, d\rho\, d\tau \frac{\left(x_i^{n-4}\right)}{x_i^{n-2}\left(y_{i-1i+1}^2 \rho^2 + \sum_{j\neq\lbrace i-1,i,i+1\rbrace}y_{ij}^2 \,x_j \right)^{n-2}}\\
&&= 2\int \frac{dx_i}{x_i} \,d\tau \int \frac{\rho \, d\rho\, \prod_{j\neq \lbrace i-1,i,i+1\rbrace}dx_i}{\left(y_{i-1i+1}^2 \rho^2 + \sum_{j\neq\lbrace i-1,i,i+1\rbrace}y_{ij}^2 \,x_j\right)^{n-2}}
\eeas
The divergent factor is 
\beas
\int \frac{dx_i}{x_i}d\tau = \int d\text{log} x_i \, d\text{log}\frac{x_{i+1}}{x_{i-1}}
\eeas
The remaining integrals are rational as expected and can be easily evaluated.
\beas
\left(\int \frac{2\rho\, d\rho \prod_{j\neq \lbrace i-1,i,i+1\rbrace} dx_i}{\left(\rho^2 +\sum_{j\neq i-1,i,i+1} x_j\right)^{n-2}}\right) = \frac{1}{(n-3)!}
\eeas
With this,
\beas
I_n && \approx \int d\text{log}\, x_i \,\,d\text{log}\left(\frac{x_{i+1}}{x_{i-1}}\right)\,\,\frac{1}{y_{i-1,i+1}^2\prod_{j\neq \lbrace i-1,i,i+1\rbrace} y_{ij}^2} \frac{1}{(n-3)!}
\eeas
We see that there is a log$^2$ divergence and its coefficient is the same as $\Gamma_{I_n}$ upto a numerical factor.
\bea
\label{eq:cusp1}
\frac{1}{(n-3)!}\frac{1}{y_{i-1i+1}^2\prod_{j\neq \lbrace i-1,i,i+1\rbrace} y_{ij}^2} = \frac{8\Gamma_{I_n}}{(n-3)!}
\eea

\subsection{Proof for general one-loop integrals}
We will now generalize the above results to include cases with tensor numerators. It is easiest to work in embedding space. A generic one-loop integral with a tensor numerator has the form 
\bea
\label{eq:embedding}
I_n = \int \frac{T\left[Y^{n-4}\right] \left[d^4 Y\right]}{(Y.Y_1)\dots (Y.Y_n)}
\eea
where $T\left[Y^{n-4}\right] = T_{i_1 \dots i_n}Y^{i_1}\dots Y^{i_n}$ is a tensor of rank $n-4$. The measure $[d^4Y] = \frac{d^6Y \delta(Y.Y)}{\text{Vol}(GL(1))}$. \\

To calculate the coefficient of the IR divergence, we follow the same procedure as in Section [\ref{sec:momsp}]. We calculate the residue on the cut $Y.Y_1 = Y.Y_2 = Y.Y_3 = 0$. Since the denominator is the same as in (\ref{eq:scalarfint}), it is easy to see that the same computation goes through. The end result is,
\bea
\label{eq:embeddingres1}
\Gamma_{I_n} =  \frac{1}{8}\frac{T\left[Y_2^{n-4}\right]}{(Y_1.Y_3)\prod_{k\neq \lbrace 1,2,3 \rbrace}Y_2.Y_k}
\eea
We will now show that the same result can be obtained in Feynman parameter space by scaling the parameters as mentioned before. We being by Feynman parametrizing the integral in (\ref{eq:embedding})
\beas
I_n = \int \frac{T\left[Y^{n-4}\right] \left[d^4 Y\right]\langle X d^{n-1}X\rangle}{(Y.W)^n}
\eeas
where $W = \sum_i x_i Y_i$. To do the integral over $Y$, we note that each factor of $Y$ can be exchanged for $\frac{d}{dW}$ to get
\beas
I_n &&= \frac{6(-1)^{n-4}}{(n-1)!} T\left[ \left(\frac{d}{dW}\right)^{n-4}\right]\int \frac{\left[d^4Y\right] \langle Xd^{n-1}X\rangle}{(Y.W)^4}\\
&&= T\left[\left( \frac{d}{dW}\right)^{n-4}\right]\int  \frac{\langle Xd^{n-1}X\rangle}{(W.W)^2} 
\eeas
where $T\left[ \left(\frac{d}{dW}\right)^{n-4}\right] = T_{i_1\dots i_{n-4}}\frac{d}{dW^{i_1}}\dots \frac{d}{dW^{i_{n-4}}}$ and we have used 
\beas
\int \frac{\left[ d^4Y\right]}{(W.Y)^4} = \frac{1}{(W.W)^2} 
\eeas
\\
To compare with (\ref{eq:embeddingres1}), we set $x_1 = \sqrt{x_2}\rho e^\tau ,x_3 = \sqrt{x_2}\rho e^{-\tau}$ and take the limit of large $x_2$. Once again we have $W \approx x_2 Y_2$ and  $W.W \approx x_2 \,\left(\rho^2 \,Y_1.Y_3 + \sum_{i\neq 1,2,3} x_i Y_i.Y_2\right)$. In the large $x_2$ limit, only the term $T[W^{n-4}] = x_2^{n-4}T[Y_2^{n-4}]$ contributes and
\beas
I_n \approx \int \frac{dx_2}{x_2}d\tau \frac{T[Y^{n-4}]}{(Y_1.Y_3)\prod_{i\neq 1,2,3}(Y_2.Y_i)}\int\frac{\langle Xd^{n-4}X\rangle \rho d\rho}{(\rho^2 +\sum_{i\neq 1,2,3} x_i)^{n-2}}
\eeas
The integral $\int\frac{\langle Xd^{n-4}X\rangle \rho d\rho}{(\rho^2 +\sum_{i\neq 1,2,3} x_i)^{n-2}} = \frac{1}{(n-3)!}$ is independent of the details of the numerator. This explains why $\Gamma_{I_n}$ is always rational at one-loop irrespective of the details of the integrand. \\

We have shown that the co-efficient of the IR divergence can be extracted from the integral by an algebraic operation directly in Feynman parameter space. There is a potential IR divergence associated with every triplet $(x_{i-1},x_i,x_{i+1})$. The complete IR divergence associated with the one-loop integral (\ref{eq:embedding}) is given by summing over all such regions 
\bea
\label{eq:gamma}
\Gamma = \sum_{i=1}^n\frac{T\left[Y_i^{n-4}\right]}{(Y_{i-1}.Y_{i+1})\prod_{k \, \neq i-1,i,i+1} Y_i.Y_k}.
\eea 
\subsection{IR Finite integrals}
It is instructive to understand what makes integrals IR finite in Feynman parameter space. We can see from $(\ref{eq:cusp})$ that $\Gamma = 0$ unless $T[Y_i^{n-4}] \neq 0$ for at least one $i \in \lbrace 1, \dots n \rbrace.$ As an example, consider a well known finite integral, the chiral hexagon
\bea
\label{eq:hexagon}
I = \int_{AB} \frac{\langle AB13 \rangle \langle AB46 \rangle \langle 5612 \rangle \langle 2345\rangle}{\langle AB12\rangle \langle AB23\rangle \langle AB34\rangle \langle AB45\rangle \langle AB56\rangle \langle AB16\rangle}
\eea
\begin{figure}[htb!]
\begin{center}
\begin{tikzpicture}
\coordinate [label=60:{\scriptsize $1$}] (e1) at (60:1.4cm);
\coordinate [label=0:{\scriptsize $2$}] (e2) at (0:1.4cm);
\coordinate [label=-60:{\scriptsize $3$}] (e3) at (-60:1.4cm);
\coordinate [label=-120:{\scriptsize $4$}] (e4) at (-120:1.4cm);
\coordinate [label=180:{\scriptsize $5$}] (e5) at (180:1.4cm);
\coordinate [label=120:{\scriptsize $6$}] (e6) at (120:1.4cm);
\coordinate (v1) at (60:.9cm);
\coordinate (v2) at (0:.9cm);
\coordinate (v3) at (-60:.9cm);
\coordinate (v4) at (-120:.9cm);
\coordinate (v5) at (180:.9cm);
\coordinate (v6) at (120:.9cm);
\draw [black,thick] (v1) -- (v2) -- (v3) -- (v4) -- (v5) -- (v6) -- cycle;
\draw [black,thick] (v1) -- (e1);
\draw [black,thick] (v2) -- (e2);
\draw [black,thick] (v3) -- (e3);
\draw [black,thick] (v4) -- (e4);
\draw [black,thick] (v5) -- (e5);
\draw [black,thick] (v6) -- (e6);
\draw [black,thick,dashed] (v1) .. controls (.2cm,0) .. (v3);
\draw [black,thick,dashed] (v4) .. controls (-.2cm,0) .. (v6);
\fill [black] (v1) circle(4pt);
\fill [black] (v2) circle(4pt);
\fill [black] (v3) circle(4pt);
\fill [black] (v4) circle(4pt);
\fill [black] (v5) circle(4pt);
\fill [black] (v6) circle(4pt);
\end{tikzpicture}
\end{center}  
\caption{Chiral Hexagon}
\end{figure}
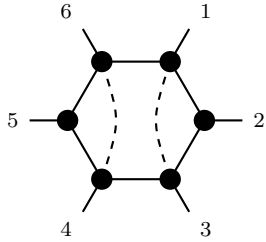\\
where we have used momentum twistor notation and $\int_{AB} = \int \langle ABd^2A\rangle \langle ABd^2B \rangle$. On Feynman parametrization, this becomes,
\beas
\frac{(Y_{13}.Y_{46}) (W.W)-6 (W.Y_{13})(W.Y_{46})}{(W.W)^4}
\eeas
where $W = \sum_{i} x_i \,Y_i$ and $Y_{ij}$ is the vector in embedding space corresponding to the bi-twistor $\vline ij\rangle$. The numerator doesn't contain any terms of the form $x_i^2$ and doesn't encounter IR divergences from the collinear region.\\

We can now easily construct a basis of IR finite integrals in Feynman parameter space. At n-points, the numerator of a Feynman integral is a polynomial of degree $(n-4)$ in the Feynman parameters. 
\bea
\int \langle Xd^{n-1}X \rangle \frac{T[X^{n-4}]}{(XQX)^{(n-2)}}
\eea 
where $T[X^{n-4}] = T_{i_1 \dots i_{n-4}}X^{i_1 \dots i_{n-4}}$. The only constraint IR finiteness imposes on T is that coefficients of $x_i^{n-4}$ should vanish for all $i \in \lbrace 1, \dots n \rbrace$.\\

At $n=5$, this implies that there are no IR finite integrals. This is in agreement with the result that the chiral pentagons for $n=5$ suffer from IR divergences from unprotected massless corners~\cite{LocalIntegrands}.\\

At $n=6$, the tensor is left with 15 independent coefficients. Further conditions can be imposed to uniquely specify a basis. For instance, we can demand that some leading singularities vanish while others are $\pm 1$. We will develop these ideas further in Section $[\ref{sec:construct}]$. But first, we need to understand the avatar of leading singularities in Feynman parameter space, which involve the notion of spherical contours.

\section{Algebraic aspects of spherical residues }
\label{sec:aspects}
The idea of a spherical contour integral and the corresponding spherical residue was introduced in \cite{SphericalContours} to compute the discontinuities of one-loop integrands directly in Feynman parameter space. Here, we give a brief description of the procedure. Consider the following integral.
\bea
\label{eq:genint}
I_{n,k} = \int \frac{\langle Xd^{n-1}X\rangle T\left[X^k\right]}{(X.Q.X)^{\frac{(n+k)}{2}}}
\eea
with $k$ and $n$ even. For any pair of Feynman parameters $(x_i, x_j)$, there is a natural decomposition of the quadric $Q$ into four parts, 
\beas
X.Q.X = && X_{\lbrace ij\rbrace}.Q_{\lbrace ij\rbrace, \lbrace ij\rbrace}. X_{\lbrace ij\rbrace}+  X_{\lbrace ij\rbrace}.Q_{\lbrace ij\rbrace, \lbrace \widehat{ij}\rbrace}. X_{\lbrace \widehat{ij}\rbrace}+\\
&& X_{\lbrace \widehat{ij}\rbrace} .Q_{\lbrace \widehat{ij}\rbrace, \lbrace ij\rbrace}X_{\lbrace ij\rbrace} + X_{\lbrace \widehat{ij}\rbrace} .  Q_{\lbrace \widehat{ij}\rbrace, \lbrace \widehat{ij}\rbrace} .X_{\lbrace \widehat{ij}\rbrace}
\eeas
where
\beas
&& X_{\lbrace ij\rbrace} = (x_i, x_j)\\
&& X_{\lbrace \widehat{ij}\rbrace} = (x_1, \dots \hat{x}_i , \dots , \hat{x}_j \dots x_n)
\eeas
and the $\hat{x}_i$ indicates that the entry is missing. The integral can develop singularities at locations determined by the entries of $Q$ (which are functions of the external momenta) and the properties of the numerator. There are possible branch point beginning at the following locations.
\beas
\bar{ij} =\begin{cases}
r(Q^{-1}_{\lbrace ij\rbrace, \lbrace ij\rbrace}) \qquad\qquad Q_{ii}\neq 0, \, Q_{jj}\neq 0\\
\left(\frac{Q_{ij}^2}{Q_{jj}}\right)^{-\text{sign}(Q_{ij})} \qquad\qquad Q_{ii} = 0, \, Q_{jj}\neq 0\\
\left(\frac{Q_{ij}^2}{Q_{ii}}\right)^{-\text{sign}(Q_{ij})} \qquad\qquad Q_{ii} \neq 0, \, Q_{jj} = 0\\
Q_{ij}^{-2\text{sign}(Q_{ij})} \qquad\qquad Q_{ii} = 0,\, Q_{jj} = 0
\end{cases}
\eeas 
These are actual branch points only if the residue on the spherical contour corresponding to the variables $x_i$ and $x_j$ is non zero. In the cases when the integral is non-zero, its value gives the discontinuity across the cut. \\

To compute the spherical residue, we use the following algorithm. \\
\begin{itemize}
\item Perform the transformation
\bea
\label{eq:transform}
\begin{pmatrix}
x_i\\
x_j
\end{pmatrix} = R\begin{pmatrix}
w_i\\
w_j
\end{pmatrix} - Q^{-1}_{\lbrace ij \rbrace \lbrace ij \rbrace}Q_{\lbrace ij \rbrace \lbrace \widehat{ij} \rbrace}X_{\lbrace \widehat{ij}\rbrace}.
\eea 
thereby reducing the denominator to the form 
\beas
&& w_i w_j + X_{\lbrace \widehat{ij}\rbrace}Q^{(ij)}X_{\lbrace \widehat{ij}\rbrace}
\eeas
Here $R$ is a 2$\times$2 matrix  such that $R^T  Q_{\lbrace ij \rbrace \lbrace ij \rbrace} R  = \begin{pmatrix}
0 && 1/2\\
1/2 && 0
\end{pmatrix}$. 
\item Integrate over the entire complex plane / Riemann sphere by setting $w_i = r e^{i \theta}$ and $w_j = r e^{-i \theta}$ with ranges $r \in (0, \infty)$ and $\theta \in (0, 2\pi)$.
\end{itemize}
It was shown in \cite{SphericalContours} that the whole procedure can be interpreted as an algebraic operation on the quadric, i.e. after integration the new quadric $Q^{(ij)}$ is related to the old one by 
\beas
Q^{(ij)} = Q_{\lbrace \widehat{ij}\rbrace \lbrace \widehat{ij}\rbrace} -  Q_{\lbrace \widehat{ij}\rbrace \lbrace ij\rbrace} Q^{-1}_{\lbrace ij\rbrace \lbrace ij\rbrace}  Q_{\lbrace ij\rbrace \lbrace \widehat{ij}\rbrace}
\eeas
Furthermore, the effect of performing multiple spherical contour integrals is captured by extensions of the same formula. In 4 spacetime dimensions, the maximum number of spherical contours we can perform is four (this is equivalent to cutting four propagators and fully localizing the momentum). This double spherical residue results in a quadric
\beas
Q^{(ijkl)} = Q_{\lbrace \widehat{ijkl}\rbrace \lbrace \widehat{ijkl}\rbrace} -  Q_{\lbrace \widehat{ijkl}\rbrace \lbrace ijkl\rbrace} Q^{-1}_{\lbrace ijkl\rbrace \lbrace ijkl\rbrace}  Q_{\lbrace ijkl\rbrace \lbrace \widehat{ijkl}\rbrace}
\eeas
In order to complete the interpretation as an algebraic operation , we need to provide similar expressions for the numerators after the integrals. We will now examine the effect the spherical contour integral has on the numerators.\\

\textbf{\underline{Linear numerator}}\\

Let's start with a Feynman parameter integral with a linear numerator, 
\bea
\label{eq:linear}
I_l = \int \frac{\langle Xd^{n-1}X\rangle (L.X)}{(X.Q.X)^{(n+1)/2}}
\eea
We want an expression for the numerator after performing a spherical contour integral along the $(x_i,x_j)$. To perform the integral, we first decompose the numerator into parts along $x_i, x_j$ and orthogonal pieces.
\beas
L.X = L_{\widehat{\lbrace ij\rbrace}}.X_{\widehat{\lbrace ij \rbrace}}+L_{\lbrace ij \rbrace}.(x_i, x_j)
\eeas  
Performing the transformation $\ref{eq:transform}$ results in an integral which we denote as
\bea
I_l^{(ij)} = \int \frac{\langle X_{\lbrace \widehat{ij}\rbrace}d^{n-3}X_{\lbrace \widehat{ij}\rbrace}\rangle (L^{(ij)}.X_{\lbrace \widehat{ij}\rbrace})}{(X_{\lbrace \widehat{ij}\rbrace}.Q^{(ij)}.X_{\lbrace \widehat{ij}\rbrace})^{(n-1)/2}}
\eea
with 
\bea
\label{eq:linearnumaftersph}
L^{(ij)} = \frac{1}{\sqrt{-4\text{Det}Q_{\lbrace ij \rbrace \lbrace ij \rbrace}}} \,\, \left(L_{\widehat{\lbrace ij\rbrace}}-L_{\lbrace ij\rbrace} Q^{-1}_{\lbrace ij \rbrace \lbrace ij\rbrace}Q_{\lbrace ij\rbrace \lbrace \widehat{ij}\rbrace}\right)
\eea

\textbf{\underline{Quadratic numerator}}\\

Consider next, the case of an integral with a quadratic numerator.
\bea
\label{eq:quadnum}
I_q = \int \frac{\langle Xd^{n-1}X\rangle (N.X.X)}{(X.Q.X)^{(n+2)/2}} 
\eea
To perform a spherical contour integral in the $(x_i,x_j)$ direction, we decompose $N^{(ij)}$ in the same way as before.
\beas
&&X.N.X = X_{\lbrace ij \rbrace} N_{\lbrace ij \rbrace \lbrace ij \rbrace} X_{\lbrace ij \rbrace}+2 N_{\lbrace ij\rbrace \lbrace \widehat{ij}\rbrace}X_{\lbrace ij \rbrace}X_{\lbrace \widehat{ij}\rbrace}+N_{\lbrace \widehat{ij}\rbrace \lbrace \widehat{ij}\rbrace}X_{\lbrace \widehat{ij}\rbrace}X_{\lbrace \widehat{ij}\rbrace} 
\eeas
We can show that the result can be written as 
\beas
I_q^{(ij)} = \int \frac{\langle X^{(ij)}d^{n-3}X^{(ij)}\rangle (X^{(ij)}.N^{(ij)}.X^{(ij)})}{(X^{(ij)}.Q^{(ij)}.X^{(ij)})^{n/2}}
\eeas
with 
\bea
\label{eq:quadratic}
\nonumber N^{(ij)} = &&Q^{(ij)} Tr(Q^{-1}_{\lbrace ij\rbrace \lbrace ij \rbrace}N_{\lbrace ij\rbrace \lbrace ij \rbrace})+(n-2)\left(Q_{\lbrace \widehat{ij}\rbrace\lbrace ij \rbrace}Q^{-1}_{\lbrace ij \rbrace \lbrace ij \rbrace}N_{\lbrace ij \rbrace \lbrace ij \rbrace}Q^{-1}_{\lbrace ij \rbrace \lbrace ij \rbrace}Q_{\lbrace ij \rbrace \lbrace \widehat{ij}\rbrace}\right.\\
 &&\left.- Q_{\lbrace\widehat{ij}\rbrace\lbrace ij \rbrace}Q^{-1}_{\lbrace ij \rbrace \lbrace ij \rbrace}N_{\lbrace ij \rbrace \lbrace \widehat{ij}\rbrace}-N_{\lbrace\widehat{ij}\rbrace\lbrace ij \rbrace}Q^{-1}_{\lbrace ij \rbrace \lbrace ij \rbrace}Q_{\lbrace ij \rbrace \lbrace \widehat{ij}\rbrace}+N_{\lbrace \widehat{ij}\rbrace \lbrace \widehat{ij}\rbrace}\right)
\eea
For more details on the calculation, we refer the reader to Appendix $\ref{sec:quadnum}$.\\

The effect of multiple spherical contours is easy to express in this form. For e.g. a double spherical residue along directions $(ijkl)$, on the linear and quadratic numerators, results in $L^{(ijkl)}$ and $N^{(ijkl)}$ with obvious definitions.
\subsection{Properties of Feynman integrals coming from loop integrals}
In this section we elaborate on some properties satisfied by Feynman integrals. An integral of the form $(\ref{eq:genint})$ must satisfy the following conditions if it comes from a Feynman diagram. 
\begin{itemize}
\item The quadric $Q$ must be degenerate for $n>6$. This is because the entries of the quadric are all of the form $Y_i.Y_j$ where $Y_i$ and $Y_j$ are embedding space vectors. The embedding space corresponding to 4D spacetime is 6 dimensional. Thus the rank of Q is always 6.
\item The tensor in the numerator, $T$ must share the null space of the degnerate Q (for $n>6$). If $N$ is a vector in the null space of $Q$, i.e. $Q.N = 0 $, then we must have $T.N =0$. \\
\end{itemize}
It is a non trivial fact that these properties continue to hold after we perform a spherical contour integral. We can use the expressions derived above to provide a quick proof of these facts. \\

This is easy to show for a Feynman parameter integral with a linear numerator (\ref{eq:linear}). We want to show that the new numerator shares a null space with the new quadric. i.e. for every $N'$ such that $Q^{(ij)}.N' = 0$, we have $L'.N=0$. To show this, suppose that $N$ belongs to the null space of $L$ and $Q$. Then we have $L.N=0 = Q.N=0$. It is easy to see that $N' = N_{\lbrace \widehat{ij}\rbrace}$ is a null vector of $Q^{(ij)}$ using the following property.
\beas
Q.N = 0 \implies Q_{\lbrace\rbrace\lbrace ij\rbrace}N_{\lbrace ij\rbrace}= - Q_{\lbrace\rbrace\lbrace\widehat{ij}\rbrace}N_{\lbrace\widehat{ij}\rbrace} \\
\eeas
where the empty $\lbrace\rbrace$ can be either $\lbrace ij\rbrace$ or $\lbrace \widehat{ij}\rbrace$.  Using (\ref{eq:linearnumaftersph}) it is obvious that $L^{(ij)}.N' = 0$. Thus (\ref{eq:linear}) satisfies all the conditions of a Feynman integral after a spherical contour.\\

This can be extended to a class of integrals of the form
\bea
\label{eq:intwithpower}
\langle Xd^{n-1}X\rangle \frac{(L.X)^{n-D}}{(X.Q.X)^{n-D/2}}
\eea
The spherical residue in variables $(x_i,x_j)$ is a sum of terms of the form $(0\leq k \leq n-D)$
\beas
\frac{(L^{(ij)}.X)^{n-D-k/2}}{(X'Q'X')^{(D+k+2-2n)/2}}
\eeas
See [\ref{sec:F}] for the detailed derivation of this result. We see that the proof for a linear numerator works here as well. An similar calculation using (\ref{eq:quadratic}) shows that the same holds true in the case of a quadratic numerator 
\subsection{Spherical contours meet IR divergences}
We have seen that the double spherical contours calculate the leading singularities. We know that leading singularities obey relations that arise from the global residue theorem \cite{LocalIntegrands}. These must be reflected in the double spherical contours. Let us start with the simple example of 
\beas
I_5=\int \frac{\langle Xd^4X \rangle\, x_2}{(x_1 x_3 Q_{13}+x_1 x_4 Q_{14}+x_2 x_4 Q_{24}+x_2 x_5 Q_{25}+x_3 x_5 Q_{35})^3}\\
\eeas
This integral is IR divergent and the divergence corresponds to the triplet $(x_1x_2x_3)$. Let us calculate the double spherical contours $(1423), (1425), (1324), (1325)$. 
\bea
c_{1423} = -c_{1425} = c_{1324} = c_{1325} = -\frac{1}{2Q_{25}Q_{13}Q_{24}}
\eea
We see that $c_{1425}+c_{1423}=0$ as expected from the Global residue theorem. However, $c_{1324}+c_{1325} \neq 0$ and this is precisely because of the IR divergence. Similar residue theorems are satisfied by the double spherical contours as can be checked from our expression for the 6 point MHV amplitude. Since the IR divergence introduces non-zero composite residues, the statement of the global residue theorem must be changed to accommodate these. The spherical residue capture the usual leading singularities in Feynman parameter space and the scaling limit introduced in Sec[$\ref{sec:feynmanres}$] captures the composite residues. A similar analysis can be found in \cite{DCregulator}. 

\section{Constructing integrands using spherical residues}
\label{sec:construct}
In $4D$, performing two spherical contour integrals is equivalent to putting four propagators on-shell. This fully localizes the loop momentum. The resulting object is the sum of the leading singularities associated with cutting the four propagators. Specifying the leading singularities (LS) puts constraints on the integrand. We can construct integrands from their singularities in Feynman parameter space using this technique. In this section, we will illustrate this with a few examples at 5 and 6 points. We will use our knowledge of the leading singularities of MHV amplitudes of $\mathcal{N}=4$ SYM to construct the one-loop integrand for the 5 and 6 point amplitudes. 

\subsection{5 point integrands}
At 5 points, a generic Feynman parameter integrand is
\beas
I_5 = \int \langle Xd^4X \rangle \, \frac{ (L.X)}{(X.Q.X)^3}
\eeas
Since we know that the only allowed poles in momentum twistor space are of the form $\langle ABii+1\rangle = 0$, we will assume that the quadric is $Q_{ij} = \langle i-1ij-1j\rangle$. The vector in the numerator $L = (l_1, l_2, l_3, l_4, l_5)$ is to be determined from the LS. We demand that all the LS are equal and for convenience, we set them equal to 1.\\

We have five unique double spherical contour integrals corresponding to the five one mass LS. We denote a double spherical residue by the four associated Feynman parameters. (Note that our Feynman parameters are labeled such that the contour $(ij)$ is equivalent to cutting propagators $\langle ABi-1i\rangle = \langle ABj-1j \rangle = 0$. The residue corresponding to $(1435)$  is
\bea
\frac{2 (l_5 Q_{13} Q_{24} + l_4 Q_{13} Q_{25} - l_3 Q_{14} Q_{25} + l_2 Q_{14} Q_{35} - l_1 Q_{24} Q_{35})}{Q_{13}Q_{24}Q_{25}}
\eea
Demanding that this be 1 imposes a constraint on the $l_i$. Similarly demanding that all the other LS are equal to one leads to the numerator
\bea
l =1/2 \left(Q_{13}Q_{14}Q_{25},\, Q_{13}Q_{24}Q_{25},\, Q_{13}Q_{24}Q_{35},\, Q_{14}Q_{24}Q_{35},\, Q_{14}Q_{25}Q_{35}\right)  
\eea
We see that the leading singularities completely determine the five point amplitude in Feynman parameter space. This should be compared with $(\ref{eq:5ptnumerator})$ which was obtained by summing all the chiral pentagons at 5 points. Note that this integrand is IR divergent and has all the divergences associated with the 5 point amplitude. \\

We can also construct an integrand with only one non zero LS. Demanding that $I_5$ has support only on the cut $(1425)$ and has unit residue results in
\beas
\int \langle Xd^4X\rangle \frac{Q_{14}Q_{25}}{4}\frac{(x_1 Q_{13}+x_5 Q_{35})}{(XQX)^3}
\eeas
It is easy to recognize that this is the Feynman parametrization of 
\beas
\int_{AB}\frac{\langle AB34\rangle}{\langle AB12 \rangle \langle AB23 \rangle\langle AB34\rangle\langle AB45 \rangle \langle AB15 \rangle}
\eeas

\subsection{6 point integrands}
A generic 6 point integrand in Feynman parameter space has a quadratic numerator.
\beas
\label{eq:6pt}
I_6 = \int \langle Xd^5X\rangle  \, \frac{X.N.X}{(X.Q.X)^4}
\eeas
$N$ is a symmetric, rank 2 tensor. The quadric $Q_{ij} = \langle i-1ij-1j\rangle $ as usual for a one-loop integral. We can always make a change of variables to reduce it to 
\beas
\label{eq:6ptquadric}
Q =
\begin{pmatrix}
0 & 0 & 1 & 1 & 1 & 0  \\
0 & 0 & 0 & u_1 & 1 & 1 \\
1 & 0 & 0 & 0 & u_2 & 1 \\
1 & u_1 & 0 & 0 & 0 & u_3 \\
1 & 1 & u_2 & 0 & 0 & 0 \\
0 & 1 & 1 & u_3 & 0 & 0
\end{pmatrix}
\eeas
We refer the reader to Appendix $\ref{sec:6pt details}$ for more details. We have three kinds of leading singularities, one-mass, two-mass easy and two-mass hard. All the two mass hard leading singularities must vanish and all the remaining ones must be equal. We normalize them to unity for convenience. For computational simplicity, we choose external data 
\beas
Z_n = \left( 1, n, n^2, n^3 \right) , \qquad n = 1, \dots 6
\eeas
The constraints on $N$ arising from specifying the leading singularities suffice to fix all but 6 of the coefficients. After implementing these constraints, the integral can be written as a sum of two terms.
\beas
I_6 = \int \langle Xd^5X\rangle \,\frac{\left(X.N_1.X + X.N_2.X\right)}{(X.Q.X)^4}
\eeas
with
\beas
X.N_1.X = &&9\left(729 x_1^2 + 810 x_1 x_2 + 81 x_2^2 + 126 x_2 x_3 + 45 x_3^2 + 50 x_3 x_4 + 5 x_4^2 - 648 x_2 x_5 \right. \\
&&\left. + 495 x_3 x_5 + 50 x_4 x_5 + 45 x_5^2 + 810 x_1 x_6 + 1215 x_3 x_6 + 207 x_4 x_6 + 126 x_5 x_6 + 81 x_6^2\right)\\
X.N_2.X =&& 2 n_{13} x_1 x_3 + 2 n_{14} x_1 x_4 + 2 n_{24} x_2 x_4 + 2 n_{15} x_1 x_5 - 2 n_{14} x_2 x_5 + 2 n_{15} x_2 x_5 + 18 n_{24} x_2 x_5 \\
&&- (10 n_{14} x_3 x_5)/9 - (10 n_{26} x_3 x_5)/9 + 2 n_{26} x_2 x_6 - 2 n_{15} x_3 x_6 - 18 n_{24} x_3 x_6 - (2 n_{13} x_4 x_6)/9 \\&&+ (2 n_{14} x_4 x_6)/9 - (2 n_{15} x_4 x_6)/9 - 2 n_{24} x_4 x_6
\eeas 
The large integers that arise in this expression are due to the choice of external data. It is tedious but possible to rewrite this expression in terms of $\langle ijkl\rangle$. The integral with numerator $X.N_2.X$ is always rational and all its double spherical residues vanish. Here, we see a clear separation in Feynman parameter space of the rational part and the transcendental part. 

\section{Feynman paramerization in planar $\mathcal{N}=4$ SYM}
\label{sec:oneloopfeynpar}
In this section, we examine the one-loop MHV integrand of $\mathcal{N}=4$ SYM. It is completely determined by its leading singularities and has a well known expression in terms of chiral pentagons. 
\bea
\label{eq:MHVamplitude}
{\mathcal{A}^{{\rm 1-loop}}_{{\rm
MHV}}=\scalebox{1.05}{{\Large$\displaystyle\sum_{\text{{\footnotesize$i\!<\!j\!<\!i$}}}\,\,\left\{\phantom{\sum_{\text{{\normalsize$i<j$}}}}\hspace{1.5cm}\right\}$}}\hspace{-3.27505cm}\raisebox{-1.05cm}{\includegraphics[scale=0.11]{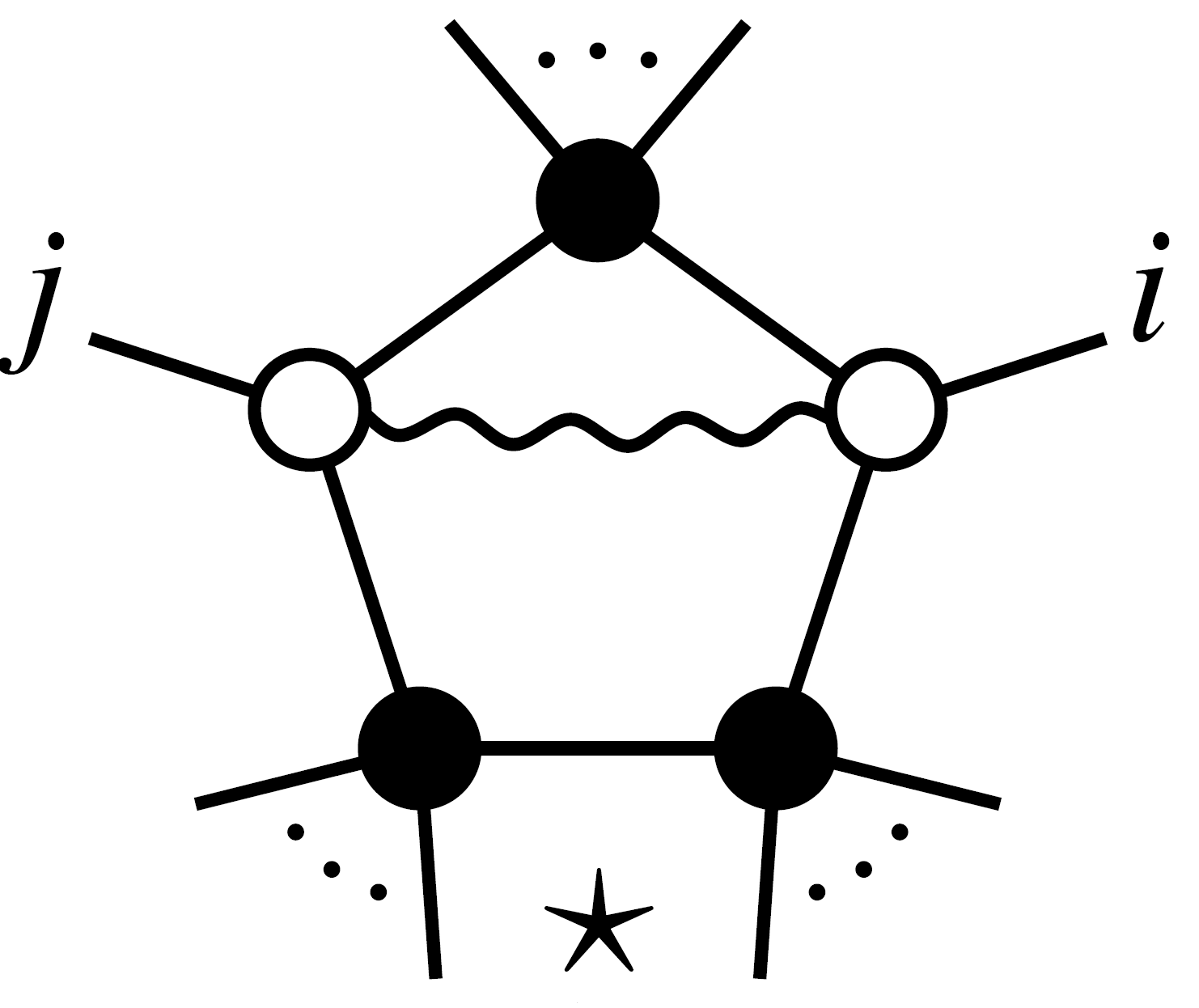}}\qquad.}
\eea
Henceforth, we denote the chiral pentagon integral shown above by $(ji)$ which takes the following form in momentum twistor space. 
\bea
\label{eq:chiralpentagon}
(ji) = \int_{AB} \frac{\langle AB (j-1jj+1)\cap(i-1ii+1)\rangle \langle \star ji\rangle}{\langle ABi-1i \rangle \langle ABii+1\rangle \langle ABj-1j\rangle \langle ABjj+1\rangle \langle AB\star\rangle}
\eea
where $\star$ is an arbitrary bitwistor.\\

There are two leading singularities, i.e. two solutions to the set of equations 
\beas
\langle ABi-1i \rangle = \langle ABii+1 \rangle = \langle ABj-1j \rangle = \langle ABjj+1 \rangle = 0
\eeas
These are the lines $Z_{\left[i\right.}Z_{\left. j\right]}$ and $(i-1ii+1)\cap (j-1jj+1)$. The above integrand is chiral and has vanishing support on the solution $(i-1ii+1)\cap (j-1jj+1)$. Thus an individual chiral pentagon is tailored to reproduce a leading singularity. However, it also has additional leading singularities arising from the pole $\langle AB\star\rangle$. These are not singularities of the amplitude and must cancel in the sum in (\ref{eq:MHVamplitude}). The cancellation of the spurious poles is not manifest and it is desirable to obtain an expression for the complete amplitude which is free of spurious poles. For attemopts along this line in momentum twistor space, see \cite{positiveamps}. Here, we will derive an expression for the complete integrand in Feynman parameter space and we will see a transparent cancellation of the spurious poles. We begin with the simple case of the four point amplitude. In this case, there are 12 contributing pentagons 
\beas
\mathcal{A}_4^{\text{MHV}}=(1,2)+(1,3)+(1,4)+ \text{cyclic} 
\eeas
\bea
\label{eq:4ptamp}
\nonumber\mathcal{A}_4^{\text{MHV}} && =\int_{AB} \frac{2\langle 1234 \rangle  }{\langle AB12 \rangle \langle AB23 \rangle \langle AB34 \rangle \langle AB41 \rangle \langle AB\star \rangle}\times 
\begin{Bmatrix}
 -\langle \star 12\rangle \langle AB34\rangle+\langle \star 23\rangle \langle AB41\rangle\\
 -\langle \star 34\rangle\langle AB12\rangle+\langle \star 41\rangle\langle AB23\rangle \\ 
 +\langle AB24\rangle\langle \star 13\rangle +\langle AB13\rangle\langle \star 24 \rangle
 \end{Bmatrix}\\
 && = \int \left[d^4Y\right] \langle Xd^4X\rangle \frac{(Y.N)}{(W.Y)^5}
\eea
where 
\beas
W = x_1 \,\vline 12\rangle + x_2 \,\vline 23\rangle + x_3 \,\vline 34\rangle + x_4 \,\vline 41\rangle + x \,\vline \star\rangle
\eeas
and $N$ is the numerator of (\ref{eq:4ptamp}) written in embedding space. Performing the momentum integral yields the Feynman parametrization.
\beas
\mathcal{A}_4^{\text{MHV}} = \int \langle Xd^4X \rangle \frac{1}{(W.W)^3}
\begin{Bmatrix}
-2\langle 1234 \rangle x(\langle \star 13\rangle\langle \star 24\rangle - \langle X12 \rangle \langle \star 34 \rangle + \langle \star 23 \rangle \langle \star 41 \rangle) \\
+ \langle 1234 \rangle^2 (x_1 \langle \star 12\rangle + x_2 \langle \star 23\rangle + x_3 \langle \star 34\rangle + x_4 \langle \star 41\rangle )
\end{Bmatrix}
\eeas
Having obtained the Feynman parametrization, it is now straightforward to demonstrate that $\mathcal{A}_4^{\text{MHV}}$ is independent of both $x$ and $\star$. First, note that the coefficient of $x$, which is quadratic in $\star$ vanishes due to a Schouten identity. The rest of the expression can be written as a total derivative.	
\bea
\label{eq:4ptfeynamp}
\mathcal{A}_4^{\text{MHV}} = -\frac{1}{2}\int \langle X d^4 X\rangle \langle 1234\rangle^2 \frac{\partial}{\partial x}\left(\frac{1}{(W.W)^2}\right)	= \frac{1}{2}\int \langle Xd^3X\rangle \frac{1}{(\tilde{W}.\tilde{W})^2}
\eea
with $\tilde{W} = W\vline_{x = 0}$ and the integral over the remaining Feynman parameters.\\

This procedure can be repeated at higher points. In each case, we find that the coefficient of the highest power of $x$ vanishes due to a Schouten identity and the rest can be writen as a total derivative which is independent of $\star$ at the boundaries. We present an expression for the 5 point amplitude. The details of the calculations are relegated to Appendix $\ref{sec:oneloopdetails}$.
\bea
\label{eq:5ptamp}
\mathcal{A}_5^{\text{MHV}}=\int \langle X d^5X \rangle  \, \frac{\partial}{\partial x} \left( \frac{2 \,n_0\, (W.\star)+\tilde{W}.\tilde{W} n_1+3 \,(W.\star) n_1 x}{(\tilde{W}.\tilde{W})^3 (W.\star)^2}\right)
\eea
Here $n_0$ and $n_1$ are the coefficients of $x^0$ and $x$ in (\ref{eq:5ptnum}). As before, the integral localizes to the boundaries where it is independent of the bitwistor $\star$ and is given by
\bea
\label{eq:5ptamplitude}
\mathcal{A}_5^{\text{MHV}} = \int \langle Xd^4X \rangle \,\frac{n}{d^3}
\eea
\bea
\label{eq:5ptnumerator}
&&\nonumber n = \left( \langle 1234\rangle\langle 1245\rangle \langle 1235\rangle x_1 + \langle 1234\rangle \langle 2345\rangle \langle 1235\rangle x_2 + \langle 1345\rangle \langle1234\rangle \langle2345\rangle x_3 \right.\\
&&\left. + \langle 1345 \rangle \langle 2345\rangle \langle 1245 \rangle x_4 + \langle 1345\rangle \langle 1245\rangle \langle 1235\rangle x_5 \right)\\\nonumber\\
&&\nonumber d = \tilde{W}.\tilde{W} \qquad \text{where  } \tilde{W} = x_1 \vline 12\rangle + x_2 \vline 23 \rangle + x_3 \vline 34 \rangle+x_4 \vline 45 \rangle + x_5 \vline 15 \rangle
\eea
It is easy to see that $(\ref{eq:4ptfeynamp})$ and $(\ref{eq:5ptnumerator})$ have the correct singularity structure. The presence of linear terms in the numerator of the 5 point amplitude implies the presence of IR divergences as expected. We can obtain similar expressions for the integrand at higher points. However, this has to be done on a case by case basis and we don't have a general expression.  \\

\section{Outlook}
In this paper we have explored the singularity structure of one-loop Feynman parameter integrands  and their geometry. The spherical residue captures the notion of discontinuity and the double spherical residue that of leading singularities. Feynman parameter integrands that arise from Feynman graphs satisfy special constraints and we saw that the spherical contour remarkably preserves these properties. We have provided an algebraic description of spherical residues and given formulae which can be use to compute both them as algebraic mappings. The double spherical residue was exploited to construct Feynman parameter integrands. Composite residues in momentum space captures the leading IR divergences. The scaling procedure introduced in Section[\ref{sec:feynmanres}] to extract the leading IR divergences shows that the notion of composite residues exists even in Feynman parameter space. \\

The obvious next step is to extend the results of this paper beyond one loop. It would be interesting to explore the extraction of the leading IR divergence of a two loop graph by a similar method. For some details on higher loop Feynman parametrization and IR divergences, we draw the reader's attention to \cite{manifestDCintegration}. While extraction of the leading IR behaviour is fascinating in its own right, it could also prove useful in calculating the cusp anomalous dimension of $\mathcal{N}=4$ SYM which has been a topic of some interest in the past few years \cite{Cusp}. The knowledge of the relationship between cuts of Feynman graphs and discontinuites is intensely studied in momentum space (see \cite{Britto1, Britto2}). In Feynman parameter space, this amounts to an underanding of the relationship between between spherical residues and leading singularities at higher loops. This is an essential ingredient in attempting any construction of higher loop integrands. While these are some of the immediate pragmatic questions of general interest, some features of Feynman parameter integrands of $\mathcal{N}=4$ SYM raise more provocative questions.\\

Section[\ref{sec:oneloopfeynpar}] shows the explicit independence of MHV amplitudes on spurious poles at 4 and 5 points. While this cancellation is expected even in momentum twistor space, it is simpler to observe in Feynman parameter space and isn't the consequence of a complicated identity satisfied by the external data. Another miraculous feature, seen from the 4 and 5 point one-loop integrands, Eqs $(\ref{eq:4ptfeynamp})$ and $(\ref{eq:5ptnumerator})$, is that they are both manifestly positive ( for positive external data). Positivity of the integrands in momentum twistor space was observed in \cite{positiveamps}. There the positvity stemmed from the more complicated identity $\langle AB \bar{i}\bar{j}\rangle >0$ for configurations of $Z_i$ in the amplituhedron. Here, $\langle ijkl \rangle >0$ for $i<j<k<l$ suffices to guarantee positivity. It is crucial to check if these features persist beyond one loop. It would also be interesting to analyze the positivity properties of the log of the amplitude and the n-point Ratio function \cite{positivity} in Feynman parameter space. \\

The existence of these properties seems to suggest that Feynman parameter space more than an auxiliary space introduces to aid in integration and is a natural space to study loop integrands. In the last decade, a rich geometric structure underlying scattering amplitudes of $\mathcal{N}=4$ SYM has been uncovered \cite{grassmannian, amplituhedron} and positive geometry  \cite{Positivegeometry} is at the heart of it all. It is a natural to wonder if the properties seen here are a reflection of this structure. If this were true, it suggests that Feynman parameter space has an extremely rich geometry and the properties observed thus far are only the tip of the iceberg.

\acknowledgments
We would like to thank Nima Arkani-Hamed for guidance at all stages of this project and for going through multiple versions of the manuscript. We also thank Ellis Yuan and Enrico Herrmann  for useful discussions. 
\appendix 
\section{Cuts of Feynman integrals}
\label{sec:F}
A class of integral coming from Feynman parametrizing a 1-loop diagram will are of the form
\beas
I_n =\int \langle Xd^{n-1}X\rangle \frac{(L.X)^{n-D}}{(X.Q.X)^{n-D/2}}
\eeas
We will perform a spherical contour integral in the $(ij)$ directions. Using the transformation in  $(\ref{eq:transform})$, the above integral becomes,
\beas
I^{(ij)}_n =&&  \int dw_i \, dw_j \, \langle X^{(ij)}d^{n-3}X^{(ij)} \rangle \mathcal{R}\, \frac{\left(L_{\lbrace\widehat{ij}\rbrace}X_{\lbrace\widehat{ij}\rbrace}+L_{\lbrace ij\rbrace}(Rw)_{\lbrace ij\rbrace}-L_{\lbrace ij \rbrace}Q^{-1}_{\lbrace ij\rbrace \lbrace ij\rbrace}Q_{\lbrace ij\rbrace \lbrace \widehat{ij}\rbrace}X_{\lbrace\widehat{ij}\rbrace}\right)^{n-D}}{{(w_i \, w_j + X^{(ij)}Q^{(ij)}X^{(ij)})^{n-D/2}}}
\eeas
where $\mathcal{R} = $det $R$. The integral over $w_i, w_j$ to be done over $S^2$ with an implicit factor of $\frac{1}{2\pi i}$. Using $\ref{eq:linearnumaftersph}$, we can write the numerator as
\beas
\left( L_{\lbrace ij\rbrace }Rw_{ij}+ L^{(ij)}X_{\lbrace \widehat{ij}\rbrace}\right)^{n-D}
\eeas
Since we are integrating over the Riemann sphere with the substitution $w_i = r e^{i \phi} , \, w_i = r e^{-i \phi}$, only terms containing some power of the product $w_i w_j$ survive the angular integration. This yields,
\beas
I_n^{(ij)} &&= \sum_{k=0,\text{even}}^{n-D} {n-D\choose k}{k\choose k/2}(R.L)_i^{k/2}(R.L)_j^{k/2}\frac{\Gamma \left(1+k/2\right)\Gamma \left(-1-D/2-k/2+n\right)}{2\Gamma \left( n-D/2\right)}\\
&& \int \langle X^{(ij)}d^{n-3}X^{(ij)} \rangle\frac{(L^{(ij)}.X^{(ij)})^{n-D-k/2}}{(X^{(ij)}Q^{(ij)}X^{(ij)})^{(D+k+2-2n)/2}}
\eeas
\section{Spherical contour with a quadratic numerator}
\label{sec:quadnum}
In this appendix, we sketch out the details of transformation of a quadratic numerator under a spherical residue. Consider the integral in $\ref{eq:quadnum}$. The transformation $(\ref{eq:transform})$ changes the numerator to 
\beas
&&X.N.X \rightarrow \text{det} R \left((Rw)N_{\lbrace ij\rbrace \lbrace ij \rbrace}(Rw)+X_{\lbrace \widehat{ij}\rbrace}Q_{\lbrace \widehat{ij}\rbrace\lbrace ij \rbrace}Q^{-1}_{\lbrace ij \rbrace \lbrace ij \rbrace}N_{\lbrace ij \rbrace \lbrace ij \rbrace}Q^{-1}_{\lbrace ij \rbrace \lbrace ij \rbrace}Q_{\lbrace ij \rbrace \lbrace \widehat{ij}\rbrace}X_{\lbrace \widehat{ij}\rbrace}\right.\\
&&\left. -2 X_{\lbrace \widehat{ij}\rbrace}Q_{\lbrace\widehat{ij}\rbrace\lbrace ij \rbrace}Q^{-1}_{\lbrace ij \rbrace \lbrace ij \rbrace}N_{\lbrace ij \rbrace \lbrace \widehat{ij}\rbrace}X_{\lbrace \widehat{ij}\rbrace}+X_{\lbrace \widehat{ij}\rbrace}N_{\lbrace \widehat{ij}\rbrace \lbrace \widehat{ij}\rbrace}X_{\lbrace \widehat{ij}\rbrace}\right) \\
\eeas
With det$ R =2\sqrt{-\text{det}Q_{\lbrace ij\rbrace\lbrace ij\rbrace}}$, we can write the cut integral as  
\beas
I_q^{(ij)} = \int \frac{\langle X^{(ij)}d^{n-3}X^{(ij)}\rangle}{2\sqrt{-\text{det}Q_{\lbrace ij\rbrace\lbrace ij\rbrace}}}\frac{dw_i dw_j}{2\pi i} \frac{(Rw)N_{\lbrace ij\rbrace \lbrace ij \rbrace}(Rw)+X_{\lbrace \widehat{ij}\rbrace}N' X_{\lbrace \widehat{ij}\rbrace}}{(w_i w_j + X^{(ij)}Q^{(ij)}X^{(ij)})^{\frac{n}{2}+1}}
\eeas
The first term integrates to
\beas
\frac{1}{2\pi i}\int_{w_i=\bar{w}_j} \frac{(w.(R^TNR).w)\, \langle X^{(ij)}d^{n-3}X^{(ij)}\rangle dw_i\, dw_j}{(w_i w_j + X_{\lbrace \widehat{ij}\rbrace}Q^{(ij)}X_{\lbrace \widehat{ij}\rbrace})^{\frac{n+2}{2}}} = \frac{ \langle X^{(ij)}d^{n-3}X^{(ij)}\rangle Tr(Q^{-1}_{\lbrace ij\rbrace \lbrace ij \rbrace}N_{\lbrace ij\rbrace \lbrace ij \rbrace})}{2\sqrt{-\text{det}Q_{\lbrace ij\rbrace\lbrace ij\rbrace}} n(n-2)\left(X_{\lbrace \widehat{ij}\rbrace}Q^{(ij)}X_{\lbrace \widehat{ij}\rbrace}\right)^{\frac{n}{2}-1}}
\eeas
and the second one to 
\beas
\frac{1}{2\pi i}\int_{w_i=\bar{w}_j} \frac{\langle X^{(ij)}d^{n-3}X^{(ij)}\rangle ( X_{\lbrace \widehat{ij}\rbrace}N' X_{\lbrace \widehat{ij}\rbrace}) dw_i\, dw_j}{(w_i w_j + X_{\lbrace \widehat{ij}\rbrace}Q^{(ij)}X_{\lbrace \widehat{ij}\rbrace})^{\frac{n+2}{2}}} =\frac{\langle X^{(ij)}d^{n-3}X^{(ij)}\rangle( X_{\lbrace \widehat{ij}\rbrace}N' X_{\lbrace \widehat{ij}\rbrace})}{2\sqrt{-\text{det}Q_{\lbrace ij\rbrace\lbrace ij\rbrace}}n( X_{\lbrace \widehat{ij}\rbrace}Q^{(ij)}X_{\lbrace \widehat{ij}\rbrace})^{\frac{n}{2}}}
\eeas
\section{Leading singularities at 6 points}
\label{sec:6pt details}
\begin{figure}[htb!]
\label{fig:boxsingularities}
\includegraphics[scale=.5]{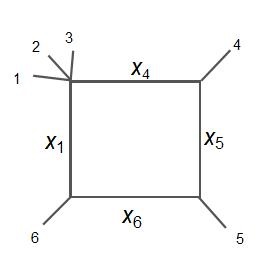}
\includegraphics[scale=.5]{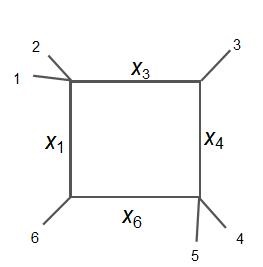}
\includegraphics[scale=.5]{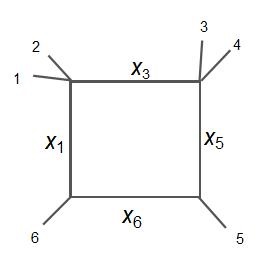}
\caption{One-mass, two-mass easy and two-mass hard singularities}
\end{figure}

At $n=6$, we can have leading singularities which correspond to the three box diagrams shown in Figure $\ref{fig:boxsingularities}$. \\
\\
We label the leading singularities by the Feynman parameters of the cut propagators. by associating $x_i$ with the propagator $\langle ABii+1\rangle$. Thus $(ijkl)$ corresponds to the leading singularity which results from setting $\langle ABi-1i \rangle = \langle ABj-1j\rangle = \langle ABk-1k \rangle = \langle ABl-1l\rangle = 0$. In this notation, the list of singularities is 
\begin{itemize}
\item {\bf One mass } (1456), (3456), (1234), (6123), (5612), (2345)
\item {\bf Two mass easy} (1356), (6245), (5134), (4623), (3512), (2461)
\item {\bf Two mass hard} (3461), (4512), (5623)
\end{itemize}
The 1-loop, $n - $ point amplitude of $\mathcal{N} = 4$ SYM is a sum over all one - mass and two-mass easy leading singularities. Thus the numerator of the full amplitude is constrained to make all the two mass hard singularities vanish and to make all the other singularities equal. The six point amplitude in momentum twistor space must have the form
\beas
\int_{AB} \frac{\langle ABX\rangle \langle ABY\rangle}{\langle AB12\rangle \langle AB23\rangle \langle AB34\rangle \langle AB45\rangle \langle AB56\rangle \langle AB16 \rangle}
\eeas
for some bitwistors X and Y. The corresponding object in Feynman parameter space looks like
\beas
\int \frac{\langle Xd^5X \rangle (X.N.X)}{(X.Q.X)^4}
\eeas
where the quadric $Q_{ij} \equiv q_{ij} = \langle i-1ij-1j\rangle$ and the numerator is a symmetric tensor with coefficients to be determined. We can simplify the denominator by making the transformation $x_i \rightarrow y_i x_i$ with
\beas
y = \left(\sqrt{\frac{q_{25}q_{36}}{q_{13}q_{15}q_{26}}},\sqrt{\frac{q_{15}q_{36}}{q_{13}q_{25}q_{26}}},\sqrt{\frac{q_{15}q_{26}}{q_{13}q_{25}q_{36}}},\sqrt{\frac{q_{13}q_{15}q_{26}}{q_{14}^2 q_{25}q_{36}}},\sqrt{\frac{q_{13}q_{26}}{q_{15}q_{25}q_{36}}},\sqrt{\frac{q_{25}q_{13}}{q_{15}q_{26}q_{36}}}\right)
\eeas
This transforms the denominator into $X.Q.X \rightarrow x_1 x_3 + x_1 x_4 + x_1 x_5 +u_1 x_2 x_4 + x_2 x_5 + x_2 x_6 +u_2 x_3 x_5 + x_3 x_6 +u_3 x_4 x_6  $ with $u_1 = q_{24} q_{15}/(q_{25} q_{14})$, $u_2 = q_{35} q_{26}/(q_{36} q_{25})$ and $u_3 = q_{46} q_{13}/(q_{14} q_{36})$.\\

We demand that all two-mass hard leading singularities vanish and that all the rest are equal to 1. This places some constraints on the numerator N. 

\section{Feynman parametrizing the MHV planar 1-loop integrand}
\label{sec:oneloopdetails}
In this appendix, we provide the details of Feynman parametrizing the complete one-loop MHV integrand for planar $\mathcal{N}=4$ SYM. As explained in Section \ref{sec:oneloopfeynpar}, the one-loop integrand is given by
\beas
\mathcal{A}_n^{\text{MHV}} = \sum_{i<j<i} (ji)
\eeas
More concretely, the expression for the amplitude at $n$ points is 
\beas
\mathcal{A}_n^{\text{MHV}} &&= \int_{AB} \left( \sum_{i=2}^n \frac{1}{\langle ABn1\rangle \langle AB12 \rangle \langle AB\star  \rangle}\frac{\langle AB (n12 \cap i-1ii+1)\rangle \langle \star 1i\rangle}{\langle ABi-1i\rangle \langle ABii+1 \rangle}\right)+\text{cyclic}
\eeas

\beas
\mathcal{A}_n^{\text{MHV}}= \int_{AB}&&\left(\langle n123 \rangle \langle \star 12\rangle\prod_{i\neq n,1,2}\langle ABii+1\rangle -\langle 12n-1n\rangle \langle \star 1n\rangle \prod_{i\neq 1,n-1,n}\langle ABii+1\rangle \right.\\
&&\left. +\sum_{k=2}^{n-1}\langle AB(n12\cap k-1kk+1)\rangle \langle \star 1k\rangle \prod_{i\neq k-1k,1,n} \langle ABii+1\rangle\ \right)\\
&&\rule{13cm}{.02cm	}\\
&&\hspace{4cm} \langle AB\star \rangle \prod_{i=1}^n \langle ABii+1 \rangle
\eeas
Combining all the terms in the cyclic sum,
\beas\mathcal{A}_n^{\text{MHV}} &&= \frac{1}{\langle AB\star \rangle \prod_{i=1}^n \langle ABii+1 \rangle}\left(2 \sum_{l=1}^n \langle \star ll+1\rangle \prod_{i\neq l-1,l,l+1}\langle ABii+1\rangle \langle l-1ll+1l+2 \rangle \right.\\
&&\left.+ \sum_{k=1}^n\sum_{l=k+2}^{k-2} \langle \star lk\rangle \langle AB(l-1ll+1)\cap (k-1kk+1)\rangle \prod_{i\neq k-1,k,l-1,l} \langle ABii+1\rangle\right)
\eeas
We can proceed with Feynman parametrization using embedding space techniques. The following formula is useful
\bea
\label{eq:FPn}
 \frac{\langle ABY_1 \rangle \dots \langle ABY_{n-3} \rangle }{\langle AB12\rangle \dots \langle ABn1 \rangle \langle AB\star \rangle} \xrightarrow[]{FP} \left(Y_1.\frac{d}{dW}\right)...\left(Y_{n-3}.\frac{d}{dW}\right)\frac{1}{(W.W)^2}
\eea
with $W = \sum_{i=1}^{n-1} x_i \,\vline ii+1\rangle + x_n \,\vline 1n\rangle + x\, \vline \star \rangle$.\\

In the 5 point case numerator after performing the cyclic sum is,
\beas
&&\langle \star 12 \rangle\langle 5123\rangle\langle AB34\rangle\langle AB45 \rangle + \langle \star 23 \rangle\langle 1234\rangle\langle AB45\rangle\langle AB51 \rangle + \langle \star 34 \rangle\langle 2345\rangle\langle AB51\rangle\langle AB12 \rangle \\
&&+ \langle \star 45 \rangle\langle 3451\rangle\langle AB12\rangle\langle AB23 \rangle + \langle \star 51 \rangle\langle 4512\rangle\langle AB23\rangle\langle AB34 \rangle  +\langle \star 13 \rangle \langle AB\bar{1}\bar{3}\rangle\langle AB45 \rangle \\
&&+\langle \star 14 \rangle \langle AB\bar{1}\bar{4}\rangle\langle AB23 \rangle +\langle \star 24 \rangle \langle AB\bar{2}\bar{4}\rangle\langle AB51 \rangle +\langle \star 25 \rangle \langle AB\bar{2}\bar{5}\rangle\langle AB34 \rangle \\
&&+\langle \star 35 \rangle \langle AB\bar{3}\bar{5}\rangle\langle AB12 \rangle .
\eeas
Eq (\ref{eq:FPn}) adapted to this case reads,
\beas
\frac{(Y_1.Y_2)(W.W)-6 (Y_1.W)(Y_2.W)}{(W.W)^3}
\eeas
which yields the following Feynman parametrization for $\mathcal{A}_5^{\text{MHV}}$.
\bea
\label{eq:5ptnum}
\nonumber &&\langle \star 12 \rangle\langle 5123 \rangle (-6\langle W34 \rangle \langle W45\rangle) + \langle \star 23 \rangle\langle 1234 \rangle (-6\langle W45 \rangle \langle W51\rangle) \\
\nonumber &&+ \langle \star 34 \rangle\langle 2345 \rangle (-6\langle W51 \rangle \langle W12\rangle)
+ \langle \star 45 \rangle\langle 3451 \rangle (-6\langle W12 \rangle \langle W23\rangle) \\
\nonumber &&+ \langle \star 51 \rangle\langle 4512 \rangle (-6\langle W23 \rangle \langle W34\rangle)\\
\nonumber &&+ \langle \star 13 \rangle \left[\langle 5124 \rangle \langle 2345 \rangle W.W - 6 \langle W45\rangle\left(\langle W12\rangle \langle 5234\rangle + \langle W25\rangle \langle 1234\rangle\right)\right]  \\
\nonumber &&+ \langle \star 14 \rangle \left[-\langle 5123 \rangle \langle 3452 \rangle W.W - 6 \langle W23\rangle\left(\langle W51\rangle \langle 2345\rangle + \langle W25\rangle \langle 1345\rangle\right)\right]  \\
\nonumber &&+ \langle \star 24 \rangle \left[\langle 1235 \rangle \langle 3451 \rangle W.W - 6 \langle W51\rangle\left(\langle W23\rangle \langle 1345\rangle + \langle W31\rangle \langle 2345\rangle\right)\right]  \\
\nonumber &&+ \langle \star 25 \rangle \left[-\langle 1234 \rangle \langle 4513 \rangle W.W - 6 \langle W34\rangle\left(\langle W12\rangle \langle 3451\rangle + \langle W31\rangle \langle 2451\rangle\right)\right]  \\
 &&+ \langle \star 35 \rangle \left[\langle 2341 \rangle \langle 4512 \rangle W.W - 6 \langle W12\rangle\left(\langle W34\rangle \langle 2451\rangle + \langle W42\rangle \langle 3451\rangle\right)\right]  
\eea
Plugging in $W$, this evaluates to $(\ref{eq:5ptamp})$

\bibliographystyle{unsrt}
\bibliography{FP}

\end{document}